\documentclass[runningheads]{llncs}

\usepackage{url}
\usepackage{graphicx}
\usepackage{graphics}

\usepackage{algorithm}
\usepackage{algorithmicx}
\usepackage{algpseudocode}
\usepackage[caption = false]{subfig}
\usepackage{color}
\usepackage{colortbl}
\usepackage{multirow}
\usepackage{array}
\usepackage{placeins}
\usepackage{adjustbox}
\usepackage{rotating}
\usepackage{hhline}
\usepackage{hyphsubst}
\usepackage{boldline}
\usepackage{soul}
\usepackage{ctable}
\usepackage{paralist}
\usepackage{amsmath}
\newcolumntype{?}{!{\vrule width 1pt}}
\makeatletter
\newcommand\footnoteref[1]{\protected@xdef\@thefnmark{\ref{#1}}\@footnotemark}
\makeatother

\newcommand{\para}[1]{\vspace{2mm}\noindent\textbf{#1}}

\newcommand{\matr}[1]{\mathbf{#1}}

\newcommand{\el}[1]{\textcolor{magenta}{\textbf{/* #1 (elex) */}}}

\begin{document}
\title{Empirical Comparison of Graph Embeddings for Trust-Based Collaborative Filtering}
\titlerunning{Graph Embeddings for Trust-Based Recommendation}
\author{Tomislav Duricic \inst{1, 2}\and
Hussain Hussain\inst{2} \and
Emanuel Lacic\inst{2} \and 
Dominik Kowald\inst{2}
\and
Denis Helic\inst{1} \and
Elisabeth Lex\inst{1}
}

\authorrunning{Tomislav Duricic et al.}

\institute{Graz University of Technology\\
\email{\{tduricic,elisabeth.lex,dhelic\}@tugraz.at}\\
\and
Know-Center GmbH\\
\email{\{hhussain,elacic,dkowald\}@know-center.at}}

\maketitle              %
\begin{abstract}
In this work, we study the utility of graph embeddings to generate latent user representations for trust-based collaborative filtering. In a cold-start setting, on three publicly available datasets, we evaluate approaches from four method families: (i) factorization-based, (ii) random walk-based, (iii) deep learning-based, and (iv) the Large-scale Information Network Embedding (LINE) approach. We find that across the four families, random-walk-based approaches consistently achieve the best accuracy. Besides, they result in highly novel and diverse recommendations. Furthermore, our results show that the use of graph embeddings in trust-based collaborative filtering significantly improves user coverage. %

\end{abstract}

\section{Introduction} 
\label{s:intro}

Recommender systems suffer from the well-known cold-start problem~\cite{schein2002methods} that arises when users have rated no or only few items. The cold-start problem is particularly problematic in neighborhood-based recommendation approaches such as collaborative filtering (CF)~\cite{schafer2007collaborative} since the ratings of these users cannot be exploited to find similar users. Trust-based recommender systems (e.g., \cite{fazeli2014,lathia2008trust,massa2004trust,donovan2005trustrecsys}) have been proposed as a potential remedy for the cold-start problem. They alleviate this problem by generating a trust network, i.e., a type of a social network in which nodes usually represent users and edges represent trust connections between users based on their explicitly expressed or implicitly derived trust relationships. Although trust is a complex and ambiguous concept from social sciences, in the context of recommender systems, we use a simple interpretation in which users trust other users in the system if they trust their opinions and ratings on different items~\cite{massa2004trust}. Resulting trust network can be used to find the $k$-most similar users, whose items are recommended to a target user. Trust networks are, however, typically sparse~\cite{kim2015sparse} since only a fraction of users have trust connections, which makes finding the $k$-similar users challenging. 
In the present work, we explore the utility of \emph{graph embeddings} to extract the $k$-similar users from trust networks. To that end, we conduct experiments on three publicly available benchmark datasets often used in studies on trust-based recommender systems: Epinions~\cite{massa2004trust}, Ciao~\cite{tang-etal12a}, and Filmtrust~\cite{guo2013novel}. We empirically evaluate a range of state-of-the-art graph embedding approaches~\cite{GOYAL2018} from four distinct method families, i.e.,
\begin{inparaenum}[(i)]
\item factorization-based methods, 
\item random-walk-based approaches, 
\item methods based on deep learning, and 
\item the LINE approach~\cite{tang2015line} that falls in neither of these families, 
\end{inparaenum}
with respect to their ability to deliver accurate, novel, and diverse recommendations~\cite{Zhang2012aurealist} for cold-start users.

In our experimental setup, we split each dataset into a validation set (warm-start users) and a test set (cold-start users). For each graph embedding approach, we perform a hyperparameter optimization on each validation set. We then select the hyperparameters which result in highest recommendation accuracy. We generate recommendations for each target user in a CF manner by finding $k$-similar neighbors using the learned embeddings and ranking their items by similarity scores. Finally, we evaluate the resulting graph embeddings against a corresponding test set with respect to accuracy and beyond accuracy metrics. We compare the graph embedding approaches against five baselines from trust-based recommender systems, commonly used in cold-start settings: (i) Most Popular (MP) recommends the most frequently rated items, (ii) $Trust_{dir}$ extracts trusted users directly from a trust network, (iii) $Trust_{undir}$ ignores edge directions and extracts neighbors from the resulting undirected network, (iv) $Trust_{jac}$ applies the Jaccard coefficient on the explicit trust network, and (v) $Trust_{Katz}$~\cite{duricic2018regular} computes the Katz similarity to infer transitive trust relationships between users. To quantify the algorithmic performance, we evaluate recommendation quality in terms of nDCG, novelty, diversity and user coverage.

We find that as a result of their ability to create a representation of each user in a network, graph embeddings are able to improve user coverage when compared to the baseline approaches. Our experiments also show that random-walk-based approaches, i.e., Node2vec and DeepWalk, consistently outperform other graph embedding methods on all three datasets in terms of recommendation accuracy. Finally, we find a positive correlation between novelty and accuracy in all three datasets suggesting that users in the respective platforms tend to prefer novel content.
Summing up, our contributions are three-fold. Firstly, we provide a large-scale empirical study on the efficacy of graph embedding approaches in trust-based recommender systems. Secondly, unlike many previous studies, which evaluated only recommendation accuracy, we compare different approaches with respect to beyond accuracy metrics such as novelty, diversity and user coverage. Lastly, our results provide new insights into user preferences based on correlations between different recommendation quality metrics.

\vspace{-2mm}

\section{Graph Embeddings} 
\label{s:approaches}

In this study, we compare the recommendation performance of graph embedding approaches from four distinct method families~\cite{GOYAL2018}, i.e., factorization-based methods, random-walk-based approaches, deep-learning-based approaches, and the LINE approach~\cite{tang2015line} which falls in neither of the first three families.

\para{Factorization-based Approaches.}
Factorization-based approaches produce node embeddings using matrix factorization. The inner product between the resulting node embedding vectors approximates a deterministic graph proximity measure~\cite{hamilton2017graphreplearning}. In total, we investigate five different factorization approaches:

\noindent - \emph{Graph Factorization (GF)}~\cite{ahmed2013distributed} factorizes the adjacency matrix and determines proximity between nodes directly on the adjacency matrix.\addtocounter{footnote}{-2}\footnote{\label{gem} Implementation used: \url{https://github.com/palash1992/GEM-Benchmark}}

\noindent - \emph{Laplacian Eigenmaps (LE)}~\cite{belkin2002laplacian} factorizes the normalized Laplacian matrix and preserves the $1^{st}$-order proximity.\footnoteref{gem} %

\noindent - \emph{Locally Linear Embedding (LLE)}~\cite{roweis2000nonlinear} minimizes the squared difference between the embedding of a node and a linear combination of its neighbors' embeddings, weighted by the edges connecting to them. The solution of this minimization problem reduces to a factorization problem.\footnoteref{gem}

\noindent - \emph{High-Order Proximity preserved Embedding (HOPE)} is able to preserve higher-order proximities and capture the asymmetric transitivity.~\cite{Ou2016}.\footnoteref{gem}

\noindent - \emph{Graph Representations with Global Structural Information (GraRep)}~\cite{cao2015grarep} can handle higher-order similarity as it considers powers of the adjacency matrix.\footnote{Implementation used: \url{https://github.com/benedekrozemberczki/role2vec}}

\para{Random Walk-based Approaches.} 
RW-based approaches first identify the context of a node with a random walk and then learn the  embeddings typically using a skip-gram model~\cite{GOYAL2018}. In total, we evaluated three different approaches:

\noindent - \emph{DeepWalk}~\cite{Perozzi2014} extracts node sequences with truncated random walks and applies a skip-gram model~\cite{Mikolov2013} with hierarchical softmax on the node pairs.\footnote{Implementation used: \url{https://github.com/phanein/deepwalk}}.

\noindent - \emph{Node2vec}~\cite{Grover2016} extends DeepWalk with hyperparameters to configure the depth and breadth of the random walks. In contrast to DeepWalk, Node2vec enables to define flexible random walks, while DeepWalk only allows unbiased random walks over the graph~\cite{hamilton2017graphreplearning}.\footnote{Implementation used \url{https://github.com/aditya-grover/node2vec}}

\noindent - \emph{Role2vec}~\cite{Ahmed2018} uses attributed random walks to learn embeddings. As Role2vec enables to define functions that map feature vectors to types, it can learn embeddings of \emph{types of nodes}.\footnote{Implementation used:  \url{https://github.com/benedekrozemberczki/role2vec}}

\para{Deep Learning-based Approaches.}
Such approaches use deep neural network models to generate node embeddings. In this research paper, we studied three deep learning-based models in total:

\noindent - \emph{Deep Neural Networks for Graph Representations (DNGR)}~\cite{cao2016deep} uses random surfing to build a normalized node co-occurrence matrix and employs a stacked denoising autoencoder to learn node embeddings.\footnote{Implementation used: \url{https://github.com/ShelsonCao/DNGR}}

\noindent - \emph{Structural Deep Network Embedding (SDNE)}~\cite{wang2016structural} finds neighbors by means of $1^{st}$ and $2^{nd}$ order proximity and learns node embeddings via autoencoders.\footnote{Implementation used: \url{https://github.com/suanrong/SDNE}}

\noindent - \emph{Graph sample and aggregate GraphSAGE}~\cite{hamilton2017inductive} is a multi-layered graph convolutional neural network, which represents nodes internally by aggregating their sampled neighborhoods and utilizes a random-walk-based cost function for unsupervised learning. GraphSAGE performs the convolution in the graph space. It uses either mean-based, GCN-based, LSTM-based, mean pooling or max pooling models for aggregation.\footnote{Implementation used:  \url{https://github.com/williamleif/GraphSAGE}}

\para{Large-Scale Information Network Embedding.}
\emph{LINE}~\cite{tang2015line} creates embeddings that preserve $1^{st}$-order and $2^{nd}$-order proximity which are represented as joint and conditional probability distributions respectively.\footnote{Implementation used: \url{https://github.com/tangjianpku/LINE}}

\section{Preliminaries}
\label{s:study}

\subsection{Datasets}
\label{s:study:datasets}
We employ three open datasets commonly used when evaluating trust-based recommender systems, i.e., Epinions~\cite{massa2004trust}, Ciao~\cite{tang-etal12a}, and FilmTrust~\cite{guo2013novel}. For all three datasets, we create an unweighted trust network, in which each node represents a user, and each directed edge denotes a trust relationship between two users. The trust network is then an adjacency matrix $\matr{A}$ where $\matr{A}_{u,v}$ is $1$ in case of a trust link between $u$ and $v$, and $0$ otherwise.
As a result of preliminary experiments, we found that most of the approaches achieved better accuracy results with an undirected network. One possible explanation is that removing the edge direction increases the average number of edges for each node and reduces the sparsity of the adjacency matrix. Moreover, some approaches are not able to consider link direction by design. Therefore, we convert the trust network to an undirected network in all of our experiments by removing edge direction, thus making $\matr{A}$ symmetric. Furthermore, we create a ratings matrix $\matr{R}$, where each non-zero entry $\matr{R}_{u,i}$ represents a rating given by a user~$u$ to an item~$i$. Table \ref{tab:dataset-stats} shows basic statistics for all three datasets.

\begin{table}[H]
\setlength{\tabcolsep}{12pt}
\centering
\renewcommand{\arraystretch}{1.25}
\caption{Dataset statistics.}
\label{tab:dataset-stats}
\scalebox{0.92}{
\begin{tabular}{l|r|r|r|r|r}
Dataset   & \#Users & \multicolumn{1}{c|}{\#Items} & \#Edges & \#Ratings & \multicolumn{1}{c}{Graph density}         \\ \hline \hline
Epinions [4]  & 49,288             & 139,738                      & 487,183             & 664,824   & $2 \times 10^{-4}$    \\ \hline
Ciao [5]      & 19,533              & 16,121                       & 40,133              & 72,665    & $1.85 \times 10^{-3}$ \\ \hline
Filmtrust [6] &  1,642                & 2,071                        & 1,853               & 35,497    & $2.43 \times 10^{-3}$
\end{tabular}
}
\renewcommand{\arraystretch}{1}
\end{table}

\para{Dataset Splits.} We split each dataset into two sets: warm-start users, i.e., users with $>10$ ratings and cold-start users, i.e., users with $\leq 10$ item ratings.
While we use the subset of warm-start users as a validation set for hyperparameter optimization concerning recommendation accuracy, the subset of cold-start users is used as a test set for measuring algorithm performance.
Table~\ref{tab:split} reports the number of cold-start and warm-start users in our datasets.

\begin{table}[t]
\setlength{\tabcolsep}{12pt}
\centering
\renewcommand{\arraystretch}{1.1}
\caption{Number of users per dataset split.}
\label{tab:split}
\scalebox{0.97}{
\begin{tabular}{c|cc|cc}
\multicolumn{1}{l|}{} & \multicolumn{2}{c|}{\textbf{Users with ratings}} & \multicolumn{2}{c}{\textbf{Users with ratings \& trust}}                          \\
\hline
   \multirow{2}{*}{Dataset}                   & \multirow{2}{*}{Warm-start} & \multirow{2}{*}{Cold-start}   & \multicolumn{1}{c}{Warm-start} & \multicolumn{1}{c}{Cold-start}  \\
& & & \multicolumn{1}{c}{(Validation set)} & \multicolumn{1}{c}{(Test set)}      \\
\hline
\multicolumn{1}{l|}{Epinions}               & 14,769 & 25,393  & 14,769 & 25,393 \\
\multicolumn{1}{l|}{Ciao}                   & 1,020  & 16,591  & 571    & 2,124  \\
\multicolumn{1}{l|}{Filmtrust}              & 963    & 545     & 499    & 241    \\
\end{tabular}
}
\renewcommand{\arraystretch}{1}
\end{table}

\subsection{Experimental Setup} 
\label{s:study:designmetrics}
The initial directed trust network is converted to an undirected network by removing edge directions. The resulting undirected symmetric $\matr{A}$ is then used as an input for the graph embedding methods, which, as a result, create a $d$-dimensional embedding for each node (i.e., user) in the graph.

\para{Recommendation Strategy.}
After generating the embedding for each node in the graph, a similarity matrix $\matr{S}$ is created based on the pairwise cosine similarity between nodes' embeddings.
Recommendations are generated in a kNN manner where we find the $k$-nearest neighbors $N_k$ (i.e., $k$ most similar users) for the target user $u_t$ using the similarity matrix $\matr{S}$. 
We use $k=40$ across all of our experiments as in~\cite{duricic2018regular}. 
Then, we assign a score for all items the users in $N_k$ have interacted with: 

\begin{equation}
\label{eq:score}
    score(i, u_t) = \sum_{v \in N_k} S_{u_t, v} \cdot R_v(i),
\end{equation}

\noindent where $R_v(i)$ corresponds to the rating assigned by the user $v$ to the item $i$ and $S_{u_t, v}$ corresponds to the similarity score in $\matr{S}$ between target user $u_t$ and the neighbor user $v$ from $N_k$.
For each target user $u_t$ with $n$ rated items, we recommend $10$ items ranked according to Eq~\ref{eq:score} and compare them with the actual rated items.

\para{Evaluation Metrics.} Previous research has shown~\cite{Kaminskas2016beyondaccuracy} that accuracy may not always be the only or the best criteria for measuring recommendation quality. Typically, there is a trade-off between accuracy, novelty, and diversity since users also like to explore novel and diverse content depending on the context. 
Therefore, in our work, we examine both novelty and diversity as well as accuracy. In particular, in our experimental setup, we use the following four accuracy and beyond-accuracy metrics for evaluation. \\
\emph{Normalized Discounted Cumulative Gain (nDCG$@n$)} -- a ranking-dependent metric measuring recommendation accuracy based on the Discounted Cumulative Gain (DCG) measure~\cite{jarvelin2008discounted}. \\
\emph{Novelty$@n$} -- corresponds to a recommender's ability to recommend long-tail items that the target user has probably not yet seen. We compute novelty using the Expected Popularity Complement (EPC) metric~\cite{vargas2011rank}. \\
\emph{Diversity$@n$} -- describes how dissimilar items are in the recommendation list. We calculate it as the average dissimilarity of all pairs of items in the recommendation list~\cite{smyth2001similarity}. More specifically, we use cosine similarity to measure the dissimilarity of two items based on doc2vec embeddings~\cite{mikolov2013distributed} learned using the item vector from $\matr{R}$ where each rating is replaced with the user id. \\
\emph{User Coverage} -- defined as the number of users for whom at least one item recommendation could have been generated divided by the total number of users in the target set~\cite{massa2004trust}.

\para{Baseline Approaches.} We evaluate the graph embeddings approaches against five different baselines: \\
 - \emph{Explicit directed trust (Trust$_{dir}$)} is a naive trust-based approach that uses the unweighted, directed trust network's adjacency matrix for finding user's nearest neighbors, i.e. $\matr{S} = \matr{A}$. \\
 - \emph{Explicit undirected trust (Trust$_{undir})$} is similar to Trust$_{dir}$ but converts the network to an undirected one by ignoring the edge direction, thus making $\matr{A}$ symmetric, i.e. $\matr{S} = \matr{A}_{undir}$. \\
 - \emph{Explicit trust with Jaccard (Trust$_{jac}$)} uses the Jaccard index on the undirected trust network $\matr{A}_{undir}$. $\matr{S}$ is a result of calculating the pairwise Jaccard index on $\matr{A}_{undir}$. The intuition behind this algorithm is that two users are treated as similar if they have adjacent users in common, i.e., trustors and trustees.  \\
 - \emph{Explicit trust with Katz similarity (Trust$_{Katz}$}) ~\cite{duricic2018regular} exploits regular equivalence, a concept from network science by using Katz similarity in order to model transitive trust relationships between users.  \\
 - \emph{Most Popular (MP)} is a non-personalized approach in recommender systems, which recommends the most frequently rated items. \\

\vspace{-8mm}

\section{Results} 
\label{s:results}

Table~\ref{tab:ndcg2} shows our results in terms of nDCG, novelty, diversity and user coverage for $n=10$ recommendations on cold-start users (test set). The reported results depict those hyperparameter configurations\footnote{Details on the hyperparameter optimization can be found at: \url{https://github.com/tduricic/trust-recommender/blob/master/docs/hyperparameter-optimization.md}}, which achieve the highest recommendation accuracy on warm users (validation set).

\subsection{Accuracy Results}

To ease the interpretation of the evaluation results across all three datasets, we rank the results by nDCG and compute the average of these ranks. Correspondingly, in the Rank$_{nDCG}$ column, we show the resulting average rank for the three datasets for recommendation accuracy.

We can observe that RW-based approaches, especially Node2vec and DeepWalk, are the best performing approaches on all three datasets.
In most cases, approaches based on graph embeddings outperform the baselines, except for Trust$_{jac}$ on Epinions. Contrary to a study conducted in~\cite{duricic2018regular}, Trust$_{jac}$ achieves higher accuracy in comparison with Trust$_{Katz}$. The reason is that in the present work, we convert the trust network to an undirected network, i.e., do not consider the direction of the trust edge. HOPE and Laplacian Eigenmaps perform best among the factorization-based approaches; LINE shows a good performance on all three datasets concerning all three metrics, and GraphSAGE is the best deep learning approach. SDNE does not perform well in our experiments, which we attribute to not exploring a sufficiently broad range of hyperparameters.

\begin{table*}[t!]
\setlength{\tabcolsep}{2pt}
\centering
\renewcommand{\arraystretch}{1.25}
\caption{\textbf{Evaluation results on cold-start users for different trust-based CF approaches for $n=10$ recommendations concerning nDCG, novelty, diversity, and user coverage comparing approaches from five different algorithm families across three different datasets. Values marked with $^{\boldsymbol{*}}$ denote that the corresponding approach was significantly better than every other approach with respect to the appropriate metric according to a Wilcoxon signed-rank test (Bonferroni corrected, $p < 0.01$). Rank$_{nDCG}$ is calculated by summing nDCG-based ranks across the datasets and re-ranking the sums.}}
\scalebox{.72}{
\begin{tabular}{c?c??c??c|c|c|c?c|c|c|c?c|c|c|c}

\multirow{2}{*}{\textbf{Cat.}} & \multirow{2}{*}{\textbf{Approach}} & %
\multirow{2}{*}{\textbf{\shortstack{Rank\\ $_{nDCG}$}}} &
\multicolumn{4}{c?}{\textbf{Epinions}} & \multicolumn{4}{c?}{\textbf{Ciao}} & \multicolumn{4}{c}{\textbf{Filmtrust}} \\ \clineB{4-15}{2.5} 
& & & \textbf{nDCG} & \textbf{Nov.} & \textbf{Div.} & \textbf{UC} & \textbf{nDCG} & \textbf{Nov.} & \textbf{Div.} & \textbf{UC} & \textbf{nDCG} & \textbf{Nov.} & \textbf{Div.} & \textbf{UC} \\
\specialrule{1pt}{0pt}{2pt} 
\specialrule{1pt}{0pt}{0pt}

\multirow{5}{*}{\rotatebox[origin=c]{90}{\textbf{Baseline}}} 
& Trust$_{dir}$ & 15 & .0245 & .0060 & .6006 & 59.2\% & .0140 & .0028 & .3700 & 3.9\% & .2655 & .0313 & \textbf{.2784} & 30.3\%   \\ \clineB{2-15}{1} 

& Trust$_{undir}$ & 15 & .0260 & .0063 & .5960 & 
97.0\%
& .0127 & \textbf{.0045} & .3632 & 
11.4\%
& .2739 & .0284 & .2731 & 
42.0\%
\\ \clineB{2-15}{1} 
& Trust$_{jac}$ & 11 & .0373 & .0056 & .6548 & 
99.9\%
& .0176 & .0027 & .3996 & 
12.8\%
& .3387 & \textbf{.0369} & .2266 & 
36.1\%
\\ \clineB{2-15}{1}
& Trust$_{Katz}$ & 12 & .0290 & .0046 & .6979 & \multirow{14}{*}{\rotatebox[origin=c]{90}{100 \%}} & .0158 & .0026 & .3842 & 
13.0\%
& .3681 & .0322 & .2185 & 
42.9\%
\\ %
\clineB{2-6}{1}
\clineB{8-15}{1}
& MP & 17 & .0134 & .0015 & \textbf{.7621}$^{\boldsymbol{*}}$ & & .0135 & .0012 & \textbf{.5666} & %
100\%
& .3551 & .0137 & .1672 & 100\% \\
\clineB{1-6}{3}
\clineB{8-15}{3}
\multirow{5}{*}{\rotatebox[origin=c]{90}{\textbf{Factorization}}} 
& LLE & 7 & .0309 & .0044 & .6977 &  & .0239 & .0036 & .4013 & 
\multirow{12}{*}{\rotatebox[origin=c]{90}{13.1 \%}}
& .3649 & .0159 & .1926 &
\multirow{12}{*}{\rotatebox[origin=c]{90}{44.2 \%}} \\ %
\clineB{2-6}{1}
\clineB{8-10}{1}
\clineB{12-14}{1}
& LE & 3 & .0318 & .0045 & .6961 &  & .0231 & .0034 & .3962 &  & .3715 & .0161 & .1853 &  \\ %
\clineB{2-6}{1}
\clineB{8-10}{1}
\clineB{12-14}{1}
& GF & 14 & .0138 & .0023 & .7024 &  & .0154 & .0022 & .3970 & & .3686 & .0154 & .1945 &  \\ %
\clineB{2-6}{1}
\clineB{8-10}{1}
\clineB{12-14}{1}
& HOPE & 3 & .0331& .0047 & .6728 &  & .0220 & .0033 & .3956 &  & .3718 & .0158 & .1827 &  \\ %
\clineB{2-6}{1}
\clineB{8-10}{1}
\clineB{12-14}{1}
& GraRep & 7 & .0298 & .0042 & .6704 &  & .0209 & .0030 & .3974 & 
& .3694 & .0147 & .1859 &  \\ 
\clineB{1-6}{3}
\clineB{8-10}{3}
\clineB{12-14}{3}

\multirow{3}{*}{\rotatebox[origin=c]{90}{ \parbox{1cm}{\centering \textbf{RW}}}} 
& Node2vec & \textbf{1} & .0413 & .0064 & .6581 &  & .0228 & .0036 & .4042 &  & \textbf{.3904} & .0151 & .2235 &  \\
\clineB{2-6}{1}
\clineB{8-10}{1}
\clineB{12-14}{1}
& DeepWalk & 2 & \textbf{.0435}$^{\boldsymbol{*}}$ & \textbf{.0067}$^{\boldsymbol{*}}$ & .6707 &  & \textbf{.0247}$^{\boldsymbol{*}}$ & .0037 & .3992 &  & .3654 & .0152 & .1950 &  \\ %
\clineB{2-6}{1}
\clineB{8-10}{1}
\clineB{12-14}{1}
& Role2vec & 6 & .0363 & .0054 & .6910 &  & .0149 & .0024 & .3933 & & .3695 & .0151 & .1919 &  \\
\clineB{1-6}{3}
\clineB{8-10}{3}
\clineB{12-14}{3}

\multirow{3}{*}{\rotatebox[origin=c]{90}{\parbox{1cm}{\centering \textbf{DL}}}} 
& DNGR & 10 & .0353 & .0051 & .6869 &  & .0197 & .0031 & .4023 & & .3583 & .0142 & .1959 &  \\ %
\clineB{2-6}{1}
\clineB{8-10}{1}
\clineB{12-14}{1}
& SDNE & 12 & .0184 & .0022 & .7412 &  & .0175 & .0028 & .3921 &  & .3687 & .0152 & .2003&  \\ %
\clineB{2-6}{1}
\clineB{8-10}{1}
\clineB{12-14}{1}
& GS & 7 & .0325 & .0047 & .6810 &  & .0216 & .0031 & .3963 &   & .3678 & .0151 & .1883 &  \\  

\clineB{1-6}{3}
\clineB{8-10}{3}
\clineB{12-14}{3}

\multirow{1}{*}{\rotatebox[origin=c]{0}{\textbf{LINE}}} 
& LINE & 5 & .0407 & .0063 & .6566 &  & .0222 & .0033 & .3992 &  & .3667 & .0150 & .1947 &  \\
\specialrule{1pt}{0pt}{0pt}
 
\end{tabular}
}
\renewcommand{\arraystretch}{1}
\label{tab:ndcg2}
\end{table*}

\subsection{Beyond-Accuracy Results}

\para{Novelty, Diversity, and User Coverage.} Being superior in the case of Rank$_{nDCG}$, Node2vec also achieves high novelty and diversity scores. Plus, it performs similarly or better than other RW-based methods across all three datasets. Factorization-based approaches show average performance concerning both novelty and diversity, except for GF, which scores very low on novelty and above average on diversity. DL approaches show average to below-average performance on novelty and average performance on diversity. Trust-based baselines achieve high novelty scores in general and, not surprisingly, MP has a high diversity score and the worst novelty score out of all approaches. Since all graph embedding approaches create a latent representation of each user in a trust network using it to generate a set of item recommendations, there are no differences among them in
user coverage. Except for MP, all baselines result in lower user coverage than the graph embedding approaches. Since MP provides the same list of recommendations to all users, it always has a maximum user coverage.

\para{Evaluation Metrics and User Preferences.} Table \ref{tab:ndcg2} reports only mean values for each of the approaches. However, we store individual nDCG, novelty, and diversity values for each target user and each approach and dataset. By computing the Kendall rank correlation coefficient (Bonferroni corrected, $p<0.01$) on non-zero metrics values for all approaches, we can get an insight into user preferences for each dataset. In this manner, we observe a statistically significant positive mean correlation across all three datasets between nDCG and novelty, ranging from $0.43$ on Epinions to $0.36$ on Filmtrust. This suggests that users on all three platforms prefer recommendations with higher novelty, especially on Epinions. We also observe a statistically significant mean negative correlation between diversity and novelty on Epinions ($-0.15$), which suggests that on this platform, more novel content seems to be less diverse.

\section{Conclusions and Future Work} 
\label{s:conclusion}

In this work, we explored the utility of graph embedding approaches from four method families to generate latent user representations for trust-based recommender systems in a cold-start setting. We found that random-walk-based approaches, (i.e., Node2vec and DeepWalk), consistently achieve the best accuracy. We additionally compared the methods concerning novelty, diversity, and user coverage. Our results showed that again, Node2vec and DeepWalk scored high on novelty and diversity. Thus, they can provide a balanced trade-off between the three evaluation metrics.
Moreover, our experiments showed that we can increase the user coverage of recommendations when we utilize graph embeddings in a $k$-nearest neighbor manner. Finally, a correlation analysis between the nDCG, novelty, diversity scores revealed that in all three datasets, users tend to prefer novel recommendations. Hence, on these datasets, recommender systems should offer a good tradeoff between accuracy and novelty. This could also explain the superior performance of the random-walk based approaches and we plan to investigate this in more detail in follow-up work.

\para{Limitations and Future Work.} One limitation of this study is that we treated the trust networks as undirected while, in reality, they are directed. This may have resulted in loss of information, and as such, we aim to further explore how to preserve different properties of trust networks (e.g., asymmetry). Moreover, it is possible that we did not examine an ample enough space of hyperparameters, which might have resulted in a more reduced performance of some of the approaches, e.g., SDNE. Furthermore, we aim to explore node properties of $k$-nearest neighbors for all methods to find and interpret the critical node properties preserved by the graph embeddings, which impact the recommendation accuracy, thus providing a better understanding of the complex notion of trust. Finally, we aim to incorporate user features obtained from the rating matrix into graph embeddings learned on the trust network.

\para{Acknowledgements.} This work is supported by the H2020 project TRUSTS (GA: 871481) and the ``DDAI'' COMET Module within the COMET – Competence Centers for Excellent Technologies Programme, funded by the Austrian Federal Ministry for Transport, Innovation and Technology (bmvit), the Austrian Federal Ministry for Digital and Economic Affairs (bmdw), the Austrian Research Promotion Agency (FFG), the province of Styria (SFG) and partners from industry and academia. The COMET Programme is managed by FFG.

\end{document}